# Endpoints for randomized controlled clinical trials for COVID-19 treatments

Lori E Dodd[1], Dean Follmann[1], Jing Wang[2], Franz Koenig[3], Lisa L. Korn[4], Christian Schoergenhofer[5], Michael Proschan[1], Sally Hunsberger[1], Tyler Bonnett[2], Mat Makowski[6], Drifa Belhadi[7,8], Yeming Wang[9,10], Bin Cao[9,10], France Mentre[7,8], Thomas Jaki[11,12]

Corresponding author: Lori E Dodd

Affiliations:
1. Biostatistics Research Branch, National Institute Allergy and Infectious Diseases, Bethesda, MD, USA.
2. Clinical Monitoring Research Program Directorate, Frederick National Laboratory for Cancer Research, Frederick, MD, USA.
3. Center for Medical Statistics, Informatics and Intelligent Systems; Medical University of Vienna, Austria.
4. Department of Medicine (Rheumatology, Allergy, and Immunology Section) and Department of Immunobiology, Yale University, New Haven, Connecticut, USA.
5. Department of Clinical Pharmacology, Medical University of Vienna, Austria
6. The Emmes Company, LLC, Rockville, MD, USA.
7. Université de Paris, IAME, INSERM, F-75018 Paris, France.
8. AP-HP, Hôpital Bichat, DEBRC, F-75018 Paris, France.
9. Department of Pulmonary and Critical Care Medicine, Center of Respiratory Medicine, National Clinical Research Center for Respiratory Diseases, Beijing, China.
10. China-Japan Friendship Hospital, Beijing, China; Department of Respiratory Medicine, Capital Medical University, Beijing, China.
11. Department of Mathematics and Statistics, Lancaster University, Lancaster, UK.
12. MRC Biostatistics Unit, University of Cambridge, Cambridge, UK.

Abstract (323 words)

**Introduction**

Endpoint choice for randomized controlled trials of treatments for novel coronavirus-induced disease (COVID-19) is complex.  A new disease brings many uncertainties, but trials must start rapidly to identify treatments that can be used as part of the outbreak response. COVID-19 presentation is heterogeneous, ranging from mild disease that improves within days to critical disease that can last weeks to over a month and can end in death.  While improvement in mortality would provide unquestionable evidence about clinical significance of a treatment, sample sizes for a study evaluating mortality are large and may be impractical, particularly given a multitude of putative therapies to evaluate.   Furthermore, patient states in between "cure" and "death" represent meaningful distinctions.   Clinical severity scores have been proposed as an alternative.  However, the appropriate summary measure for severity scores has been the subject of debate, particularly in the context of much uncertainty about the time-course of COVID-19.  Outcomes measured at fixed time-points, such as a test comparing severity scores between treatment and control at day 14, may risk missing the time of clinical benefit.   An endpoint such as time-to-improvement (or recovery), avoids the timing problem. However, some have argued that power losses will result from reducing the ordinal scale to a binary state of "recovered" vs "not recovered."

**Methods**

We evaluate statistical power for possible trial endpoints for COVID-19 treatment trials using simulation models and data from two recent COVID-19 treatment trials.

**Results**

Power for fixed-time point methods depends heavily on the time selected for evaluation.   Time-to-improvement (or recovery) analyses do not specify a time-point. Time-to-event approaches have reasonable statistical power, even when compared to a fixed time-point method evaluated at the optimal time.

**Discussion**

Time-to-event analyses methods have advantages in the COVID-19 setting, unless the optimal time for evaluating treatment effect is known in advance. Even when the optimal time is known, a time-to-event approach may increase power for interim analyses.

**Introduction**

Designing clinical trials for treatments for novel infectious disease brings many challenges, especially during a rapidly evolving pandemic. A new disease brings uncertainties arising from an imperfect understanding about the illness, little information about putative treatments, and complexities in measuring relevant patient outcomes. A pandemic adds an overloaded medical system with limited resources for research, heightened pressure to find effective treatments quickly, and unpredictability about potential case numbers. Studies need to start quickly for enrollments to track the epidemic curve. However, early on, information about endpoints may be lacking. This means trial design should be appropriately flexible to respond to new information, but without compromising scientific rigor.

COVID-19 has a heterogeneous presentation and clinical course, ranging from asymptomatic to critical disease (Table 1).[1] While most infected patients present with asymptomatic or mild disease, some develop severe or critical illness that can result in acute respiratory distress syndrome and death. The most common symptoms are fever, dry cough, dyspnea, chest pain, fatigue and myalgia, while less common symptoms are headache, dizziness, abdominal pain, diarrhea, nausea and vomiting. Most patients present with signs of bilateral pneumonia[2]. Neurologic symptoms including taste and smell disorders have been reported, with rare case reports of severe central nervous system affections.[3] Thrombotic complications in critically ill patients have also been observed.[4] Importantly, some COVID-19 patients recover quickly with limited (or no) complications, while patients suffering from severe disease may take 6-8 weeks or longer for full recovery.[5] This broad range of disease severity makes finding a common endpoint for all COVID-19 trials impractical. Endpoints for a study population representing a broad spectrum of disease may be different than those for a study with a narrow spectrum of disease.

We describe key considerations for selecting endpoints for COVID-19 treatment trials. We evaluate endpoints according to clinical relevance, ease and reliability of measurement, interpretability of its associated statistical analysis, and statistical efficiency. We discuss differences between fixed time-point endpoints and those that naturally incorporate changes over time. We evaluate statistical efficiency of multiple approaches with simulation models, as well as using data from two published COVID-19 randomized trials.[6,7]

**Methods**

*Endpoint selection*

Treatments for COVID-19 are intended to be curative, with the goal that the patient will survive and ultimately return to normal function. This contrasts with a disease such as stroke in which the goal of a treatment may be to reduce stroke-induced impairments that occur across a spectrum.[8] Likewise, a benefit on mortality would be strong evidence of an effect, but deaths are relatively rare. A study powered for mortality benefit would require a large sample size. For example, a sample size of around 2,000 is needed (for a two-arm study) to detect a hazard ratio (for death) of 0.65 with 85% power and a type I error rate of 5% with a 10% mortality rate. Lower mortality rates require even larger studies. In a setting with multiple putative therapies, studies powered for mortality will restrict the number of therapies evaluated, which may slow provision of effective treatments to support the outbreak response.

Furthermore, multiple clinical states in between "death" and "cure" represent meaningful patient states. The World Health Organization (WHO) proposed an ordinal scale ranging from death to full health, with states in between corresponding to the need for hospitalization, oxygen support (including type of support needed), and need for additional medical support (Table 1).[9] These states are important markers of how a patient feels and of disease progression (or improvement). Mechanical ventilation (intubation) marks a considerable worsening, as intubated patients often require treatment with sedatives and even paralytics to address patient discomfort and maximize therapy. Intubation is also associated with a host of complications leading to additional mortality and morbidity, such as ventilator-associated pneumonia[10], GI bleeding[11], and severe physical deconditioning. In a case series of 5,700 COVID-19 patients in New York, considerable numbers of patients remained intubated during the entire study[12]. Shortening the duration in a state like intubation or avoiding intubation altogether is of direct clinical benefit.

Timing of endpoint evaluation is another important consideration. A treatment effect that occurs early but dissipates over time may not be clinically meaningful. A treatment effect may be missed if evaluation is too early, before an intervention has had time for an effect. Timing of measurement is therefore crucial and can be particularly challenging in a novel disease with substantial heterogeneity. Time-to-event endpoints do not require specifying a fixed time (just the observation interval) and are more robust in this regard. We note that longitudinal models of other endpoints are possible, such as a mixed-effects proportional odds model[13] but are not commonly used.

Table 2 describes multiple endpoints considered for COVID-19, largely from the perspective of a definitive (Phase 3) trial. Endpoints for earlier phase studies may focus on evaluating mechanism (e.g., targeting a specific pathway) or evaluating activity so that "go/no-go" decisions for further evaluation in larger trials can be made. Endpoints are evaluated according to ease of measurement, reproducibility, whether they are clinically meaningful, and their ability to capture multiple clinical states and the time-course of disease.

Meaningfulness and reproducibility can be distorted when states are influenced by external factors, as may happen when patient numbers exceed hospital capacity. For example, ordinal categories become less meaningful when mechanical ventilators are not available and patients who would normally be in this category are shifted to others (or when guidelines recommending early intubation are followed more rigorously in some centers than others). Further, non-invasive ventilators or high-flow oxygen devices may not be utilized in settings where personal protection equipment is limited (or in the absence of negative pressure rooms) due to concerns about health-care worker infection from viral aerosolization. Similarly, hospitals exceeding capacity may discharge patients early due to demand for beds. Additional concerns have been raised that one-unit changes in the ordinal scale are not equally important. For example, extubation may represent a more meaningful improvement than being moved from high-flow oxygen to standard, low-flow oxygen. Both improvements have implications on health system resources; however, from the patient view, they may not be equal.

Endpoints used in other diseases have been considered. For example, the National Early Warning Score (NEWS2)[14] captures clinical deterioration in patients, but is not specific to COVID-19 and might not be sensitive enough for this disease. Other measures, such as SOFA[15] are well validated but are specific to ICU patients. Patients who require intensive care have a high mortality of approximately 30 to 60%.[16, 17, 18]

Multiple laboratory parameters are associated with deterioration of clinical status, including surrogates for organ injury and markers of systemic inflammation, e.g., markers of cardiac injury (troponin T), elevated liver transaminases, creatinine levels, procalcitonin levels, D-Dimer concentrations, fibrinogen[19], lactate dehydrogenase[20], and lymphopenia.[21] Elevations in C-reactive protein (CRP) and ferritin, further reflective of high levels of systemic inflammation, are also associated with severe disease, consistent with the observed hyperinflammatory syndrome that appears to occur in a subset of patients.[21] While tracking these parameters is important to better understand COVID-19, they do not directly measure how a patient functions or feels and may not correlate with clinical outcome. In supplementary Table S1, we provide examples of endpoint choices for several COVID clinical trials.

*Statistical Considerations*

To evaluate statistical considerations in more depth, we focus on four outcomes: time to death, time to recovery/improvement, ordinal scale at a fixed time point, and ordinal scale averaged across time points. We note that, with time-to-improvement/recovery models, the competing event of death requires special handling. Patients who die during follow-up should not be censored at time of death, as that assumes their recovery time would be like all who remain alive and unrecovered at that time. To state the obvious, once dead, a patient cannot recover. A death must be set to an infinite recovery time, so that at the end of follow-up, the patient is counted as "not recovered." We achieve the same objective by censoring deaths at the last observation day. Therefore, patients censored on the last observation day reflect two different states: death and failure to recover by day 28. Standard survival analysis methods can then be applied, but the "hazard" ratio refers to the instantaneous risk of a good outcome. Hence, we use the term "recovery rate ratio" (or "improvement rate ratio"). We note that, with administrative censoring from staggered entry before day 28, this approach corresponds to the Fine-Gray approach to competing risks.[22] With staggered entry, Fine-Gray censors deaths at the time they would have been censored had they not died (i.e., time of administrative censoring).

Discretizing a continuous variable is commonly thought to result in a loss of efficiency.[23,24] Similarly, reductions in efficiency may occur when an ordinal scale is discretized into a binary endpoint and others have emphasized power advantages of a proportional odds model.[25,26] Graubard and Korn note that rank-based methods (such as the proportional odds model) may have lower power when the marginal sums are not nearly uniform, compared to methods that use pre-assigned numeric values (scores) for categories of the ordinal scale.[27] Nonetheless, collapsing information can sometimes increase power. For example, if the distribution of a continuous endpoint is skewed or has wide tails, rank-based methods, or even dichotomizing and using a test of proportions, can be more powerful than a t-test. Relatedly if assignment to some ordinal categories is haphazard, methods that collapse categories can provide more power. Dichotomizing can also be useful when there is a clear cut-point beyond which negative sequelae of a disease manifest, such as with hemoglobin A1c or fasting glucose in diabetes. Table S2 provides a description of many statistical analysis options.

The endpoints considered are difficult to compare theoretically with respect to power. For example, time to recovery dichotomizes an ordinal scale into "recovered" and "not recovered", so one might assume there should be a loss in power associated with using this approach. However, time to recovery incorporates health states on multiple days instead of just one, which can increase power. For instance, if the proportional odds model is evaluated so early that no one has recovered (or so late that everyone has recovered), power for the proportional odds model on that day will be very low. Using an analysis that incorporates the average ordinal score over multiple days solves that problem, but its power gain is not as great as one might imagine because measurements on the same individual on different days are likely to be highly correlated. Furthermore, a between-arm difference in an average score may also be more difficult to interpret. For example, what does an average improvement of 0.4 units on an ordinal scale mean?

We also note that time-to-event analysis is advantageous from the perspective of interim analyses, as data from all patients with any amount of follow-up time are included. This contrasts with a fixed timepoint analysis, which only includes observations from patients who have made it to the prescribed follow-up milestone (e.g., all 14 days). In rapidly enrolling trials, time-to-event analysis may improve power to evaluate early efficacy (or harm) of treatments, and hence increase the speed at which treatment recommendations can be made.

*Evaluation of statistical efficiency and interpretability of methods*

Power is compared using two simulation methods and applications to two published studies of COVID treatments. For the simulation studies, ordinal trajectories were generated according to a random line, $\theta_{0i} + \theta_{1i} \log(d)$ for person i, where d is the day since randomization. For day d, the ordinal score for that day was given as floor[$\theta_{0i} + \theta_{1i} \log(d)$], where the notation floor[x] indicates the integer part of x. Death (score=7) and recovery (score=1) were considered absorbing states (i.e., values above 7 or below 1 were set to 7 and 1, respectively). One can visualize the trajectory as a subject deterministically sliding up or down their own "line of destiny" over 28 days and reporting their integer value each day. Loosely, 10% (5%) of placebo (active) patients were destined to die (having a large value of $\theta_{1i}$) within the 28-day observation period. The remaining subjects were destined to recover (with negative value of $\theta_{1i}$). Multiple parameter values for generating $\theta_{0i}$ and $\theta_{1i}$ were considered until trajectories roughly reflected our understanding of COVID-19 disease progression. Figure 1 depicts results for the reference scenario. Each setting was simulated 1,000 times, with 800 subjects total, equal randomization to the two arms, and 28 days of follow-up. We evaluated the proportional odds model at different days, a Wilcoxon rank-sum test on the mean ordinal score (1-7) up to day 28, a test of proportions on day 28 mortality, and Cox models for time to (a) recovery, (b) a 2-point improvement, and (c) death. One possible criticism of the above simulations is that the proportional odds assumption may not hold. A second set of simulations compared methods under the proportional odds assumption. Technical details and results are given in the appendix.

Patient-level data from two published studies were obtained to compare methods. The Adaptive COVID-19 Treatment Trial stage 1 (ACTT-1) randomized 1,062 patients to remdesivir or placebo and followed patients for 28 days.[6] The primary outcome was time-to-recovery, although ordinal scales were also assessed. Due to a surge in enrollments, the study exceeded its target sample size of 400 recoveries, reaching 482 by the time of the planned DSMB interim analysis. Data were taken from a preliminary report from an April 28, 2020 data freeze (before results were made public and before actively enrolled patients were offered cross-over treatment). Data cleaning for this data snapshot are ongoing, and the results presented here are intended to inform trial design. We compare empirical power for various methods with repeated random sampling of 50, 150 and 300 per arm. For each sample size, we replicated random sampling 100,000 times. Additionally, we present multiple analyses applied to the LOTUS study of lopinavir/ritonavir by Cao et al.[7] This study was stopped prior to reaching the pre-planned sample size. We present analyses with the original data (199 patients) as well as with hypothetical augmented data corresponding to 398 patients.

**Results**

*Simulation studies*

Power comparisons for simulations are shown in Table 3. For the reference scenario, the proportional odds model has increasingly better power for later days, with highest power at day 28. Empirical power for both time-to-(2-point) improvement and time-to-recovery is somewhat lower than that for the proportional odds model at the optimal time. Empirical power for mortality is notably lower than for other methods, which is no surprise due to the low event rate and modest effect. We explored four perturbations from this reference scenario to more fully assess performance. The perturbations were 1) lagged treatment effect, 2) faster recovery, 3) faster mortality and 4) effect solely on mortality. (Table S4). Under the lagged effect scenario, power for the proportional odds model decreases at days 7 and 14 but is similar on day 28 (compared to the reference scenario). This underscores the fragility in getting the day right with the proportional odds model. The faster recovery scenario has similar relative behavior to the reference scenario though power is uniformly increased. The faster mortality scenario has power like the reference scenario. These two perturbations show some robustness of the conclusions of the reference scenario. The last row of Table 3 provides scenarios with differences between arms from mortality only. Here, mortality has the highest power, as expected. More deaths on placebo necessarily implies more recoveries on treatment, which is why power for both time to improvement and time to recovery is around 30%.

Simulation studies under models that enforce the proportional odds assumption are provided in Table S3 and Figure S1. Results from these simulations show are similar. Namely, when the fixed timepoint is chosen well, the proportional odds model performs well but suffers a loss of power if the time point is chosen poorly.

*Applications to published COVID-19 treatment studies*

Table 4 shows estimates and p-values from various models applied to the ACTT-1 study data. At the time of the data snapshot, the following proportion of subjects had ordinal score data available: 91% day 7; 89% day 14, 74% day 21 and 70% day 28. On the observed data, the proportional odds model estimates decrease over time (opposite the simulation results above), with estimates of 1.62, 1.50, 1.42, 1.34 for days 7, 14, 21 and 28, respectively. The odds ratio of 1.50 at day 14 indicates a 50% increase in the odds of a one-category improvement for remdesivir relative to control (at day 14). The test of mean difference between arms at days 7, 14, 21 and 28 gives estimates of 0.56, 0.62, 0.53, 0.41 for days 7, 14, 21, and 28, respectively. The average difference at day 14 indicates an average improvement of 0.62 on the ordinal scale for remdesivir relative to placebo. The mean difference of the time-average (days 7, 14, 21 and 28) was 0.56. The time to recovery and time to (1- and 2-point) improvement analyses give estimates of 1.32, 1.29 and 1.28, respectively. The recovery rate ratio of 1.32 indicates a 32% faster (instantaneous) rate of recovery with remdesivir (relative to placebo). The hazard ratio (for mortality) of 0.70 indicates a lower hazard of death in the remdesivir group.

Table 4 also shows empirical power (proportion of statistically significant p-values <0.05 out of the 100,000 simulations) for sample sizes of 50, 150, and 300 per group. Power is greatest at day 7 using the proportional odds model, with rejection rates of 24%, 62% and 97% for sample sizes of 50, 150 and 300 per group. Results for the t-test were similar, with rejection rates of 22%, 59% and 95% for the three samples sizes. Rejection rates for the proportional odds model and t-test evaluated at day 14 were lower for all sample sizes (day 14 proportional odds rejection rates: 16%, 41% and 79%; day 14 t-test rejection rates: 17%, 46% and 85%, respectively for sample sizes of 50, 150 and 300 per group). By day 28, empirical power was lower, although the t-test rejection rates were higher than for the proportional odds (proportional odds rejection rates: 7%, 13% and 19%; t-test rejection rates: 9%, 20% and 40%, respectively for sample sizes of 50, 150 and 300 per group). The lower number of observations at the later time point explains some of the loss in power, although not entirely. The proportion with observations at day 7 and day 14 was similar (91% vs 89%), and the power reductions were considerable (62% vs 41% for the proportional odds model at day 7 and 14, respectively, with 150 per group).

Rejection rates for the recovery rate ratio were 18%, 48% and 87%, respectively for sample sizes of 50, 150 and 300. Results for the time to improvements were 19%, 51% and 90% (one-point improvement) and were 17%, 44% and 84% (two-point improvement) for sample sizes of 50, 150 and 300 per group, respectively. Rejection rates for the hazard ratio for mortality were 7%, 12% and 18%, for the three sample sizes considered, consistent with the low power for mortality in this setting.

Table S3 in the appendix show results from the LOTUS study of lopinavir/ritonavir. In the observed and augmented data analysis, none of the days the proportional odds was estimated were statistically significant, while with the augmented data, the time to a two-point improvement indicated a 31% faster rate of improvement with p<0.05.

**Discussion**

One important challenge with COVID-19 is disease heterogeneity.  An endpoint of cure or death would be the strongest clinical evidence of treatment effect. Trials using these endpoints may take an unfeasibly long time and preclude evaluation of other candidate treatments. The WHO ordinal scale reflects meaningful patient states. However, distinctions between categories may depend on limited resources (such ventilators or high-flow oxygen devices).  Further, local differences in standard of care (including different guidelines recommending early intubation and/or limiting non-invasive oxygen treatments) may affect results in multicenter trials. Ideally, such guidelines would be unified within clinical trials, but dogmatic restrictions could limit enrollments.  A placebo-controlled trial will reduce the potential for subjectivity to influence changes made to a patient's status.

Studies need to be launched quickly in order to inform the response, at a time when little information about the disease may be available. Planning for additional trial flexibility, without compromising scientific rigor, is important.[28]  Changes made to endpoints based on results external to the trial (and prior to reviewing data) are acceptable.[29]  In the ACTT-1 trial, the initial primary endpoint was the proportional odds model at day 14, based on early WHO guidance that recommended an analysis of ordinal scale at a fixed time point.   At the time, many thought the clinical course was more like influenza illness, with recoveries occurring over two weeks. However, in late February, it became apparent, that the course of illness was more prolonged than previously thought.  Consequently, follow-up was extended to 28-days and simulation results (presented in this manuscript), revealed the fragility of a fixed-time point analysis and highlighted the advantages of a time-to-recovery endpoint.

While both simulations and our examples show that power is comparable between a fixed-time point analysis and a time-to event analysis if the timing of the former is chosen well, marked power losses are apparent when this is not the case. Additionally, we believe that time-to-improvement/recovery analysis is easier to interpret. We also note that improvement in time-to-improvement/recovery is of relevance to the patient, as an indicator of faster improvement in clinical status, but also to a health system at maximum capacity. While a mortality improvement would have provided stronger evidence about treatment efficacy, initial estimates indicated a sample size of about 2,000 would be needed. This was deemed impractical given the goal to evaluate multiple therapeutic candidates.

The time-to event analysis offers other advantages such as that, for interim analyses, all data collected up until the data freeze were included, which can be important in an outbreak setting with rapid study enrollment. The PALM Ebola virus disease treatment trial provides one example.[30] In PALM, the primary endpoint was 28-day mortality. Due to rapid enrollment, there was a striking discrepancy between the number enrolled and the number with 28 days of follow-up. At the August 9, 2019 Data and Safety Monitoring Board meeting, 673 patients (of the 725 target) were enrolled but only 376 had 28-day follow-up; the study had enrolled 93% of its targeted sample size, but information (for the mortality proportion at day 28) was only 52%. A time-to-event analysis would have included data on all participants (for their observed time), and information would have been 65-70% at this analysis.

In our evaluations of the ACTT-1 data, day 7 had the highest power. However, evidence of an effect this early would likely not have been convincing for a definitive trial. A day 7 evaluation may be more appropriate for phase 2 trials. An alternative to the time-to-event approach would have been to specify multiple outcomes (e.g., ordinal scale at day 7, 14, 21 and 28), with multiplicity adjustments. This was considered but concerns were raised about interpretation and the need to focus on an important measure of clinical benefit.

Regardless of the primary endpoint chosen, collection of core outcome measures will ensure comparability across studies and will be important for subsequent efforts to synthesize data from different trials.[31]

Table 1. NIAID disease severity categories and WHO ordinal scale

| NIAID disease severity categories | WHO Ordinal Scale |
|---|---|
| **Asymptomatic/Presymptomatic infection:** Individuals who test positive for SARS-CoV-2 but have no symptoms | 0- Uninfected, no clinical or virological evidence of infection |
| **Mild illness:** Individuals who have any of various signs and symptoms (e.g., fever, cough, sore throat, malaise, headache, muscle pain) without shortness of breath, dyspnea, or abnormal imaging | 1- Ambulatory, no limitation on activities |
| **Moderate illness:** Individuals who have evidence of lower respiratory disease by clinical assessment or imaging and a saturation of oxygen ($SaO_2$) >93% on room air at sea level | 2- Ambulatory, limitation on activities |
| **Severe Illness:** Individuals who have respiratory frequency >30 breaths per minute, $SaO_2$ ≤93% on room air at sea level, ratio of arterial partial pressure of oxygen to fraction of inspired oxygen ($PaO_2/FiO_2$) <300, or lung infiltrates >50% | 3- Hospitalized, mild disease, no oxygen therapy |
| **Critical Illness:** Individuals who have respiratory failure, septic shock, and/or multiple organ dysfunction | 4- Hospitalized, mild disease, oxygen by mask or nasal prongs |
| | 5- Hospitalized, severe disease, non-invasive ventilation or high-flow oxygen |
| | 6- Hospitalized, severe disease, intubation and mechanical ventilation |
| | 7- Hospitalized, severe disease, ventilation and additional organ support—pressors, RRT, ECMO |
| | 8- Death |

Table 2. Possible endpoints for trials in COVID-19, corresponding target population, categorization of whether the endpoint is clinically meaningful, captures the diverse nature of disease, easy to measure and reproducible.

| Endpoint | Example | Population | Clinically meaningful | Multiple disease states | Time element | Easily measurable | Reproducibility | Additional comments |
|---|---|---|---|---|---|---|---|---|
| **Binary outcomes** | | | | | | | | |
| Mortality | Death by 28 | Moderate Severe Critical | + | o | o | + | + | + Most relevant in severe/critical disease.<br>− May miss other meaningful improvements in patient status.<br>− Requires large sample size |
| Recovery(discharge, discharge-eligible) | Recovered by day 28 | Moderate Severe | + | o | o | + | o | − May require long observation times in higher severity populations.<br>− Deaths require special consideration |
| Respiratory Failure | ECMO or mechanical ventilation | Moderate Severe | + | o | o | + | o | − Depends on resources<br>− Deaths require special consideration |
| Hospitalization | Admission within 28 days | Mild | + | − | o | + | o | − Depends on resources<br>− Does not capture improvement<br>− Deaths require special consideration |
| ICU admission | Admission within 28 days | Moderate | + | − | o | + | o | − Depends on resources<br>− Does not capture improvement<br>− Deaths require special consideration |
| **Ordinal outcomes** | | | | | | | | |
| Ordinal disease severity scale | WHO scale at a fixed day | Moderate Severe | + | + | − | o | o | − Depends on resources<br>− Defining clinical benefit less straightforward |
| **Time-to-event outcomes** | | | | | | | | |
| Time-to-recovery | Time to discharge or eligible for discharge | Moderate Severe | + | o | + | + | o | − Depends on resources<br>− Potential for "relapse" (sustained improvement removes this concern)<br>− Deaths require special consideration |
| Time to 1- or 2-point improvement in ordinal scale[1] | Time to 2-point improvement in WHO ordinal scale | Moderate Severe Critical | + | o | + | o | + | − Changes in categories must be meaningful and should be considered equally important |
| | | | | | | | | − Potential for "relapse" (sustained improvement removes this concern) |
| Time to intubation or death | | Moderate Severe | + | - | + | + | o | |
| **Continuous outcomes** | | | | | | | | |
| National Early Warning Score (NEWS score) | | Moderate Severe | o | + | o | - | + | + Familiar measure<br>− Not disease specific and hence not as sensitive to certain aspects of COVID<br>− Deaths need special consideration |
| Viral load /Viral Clearance | | Mild Moderate Severe Critical | - | o | o | - | - | − Difficult to reliably measure<br>− Relation to clinical outcomes not well established<br>− Deaths need special consideration |
| Oxygen, SpO2/FiO2 or paO2/FiO2 | Daily SpO2/FiO2 until discharge, death or 28 days | Mild Moderate | o | o | o | - | + | − Relation to clinical outcomes not well established.<br>− Modified by oxygen supplementation<br>− SpO2/FiO2 not well-validated<br>− paO2/FiO2 only broadly available for ICU patients<br>− Deaths need special consideration |
| Duration of a specific ordinal state | Hopitalization days; Mechanical ventilation days | Severe Critical | o | - | o | + | o | + Captures dimension meaningful to health system<br>+ Depends on the resources available<br>+ Deaths need special consideration |
| FLU-PRO | Change from baseline to day 14 | Mild Moderate Critical | o | + | - | o | o | + Captures aspects important to patients<br>− Deaths need special consideration<br>− Not validated for COVID-19 |
| SOFA score | Change from baseline to day 14 | Severe Critical | o | + | - | o | + | + Captures disease severity and incorporates most relevant organ systems<br>− Familiar for ICU setting<br>− Not validated for COVID-19 and not disease-specific<br>− Deaths need special consideration |

"+" indicates good performance, " −" indicates poor performance on this characteristic, neutral is denoted by "o".

Table 3.  Simulated power for different analysis methods under various scenarios for simulations (type 1 error rate = 5%).

| Scenario | Proportional Odds | | | | Mean Score | Time-to-event | | | Proportion 28 Day Mortality |
| --- | --- | --- | --- | --- | --- | --- | --- | --- | --- |
| | Day 1 | Day 7 | Day 14 | Day 28 | | Time to 2-point improvement | Time to Recovery | Time to Death | |
| Reference | 0.05 | 0.76 | 0.85 | 0.88 | 0.80 | 0.81 | 0.82 | 0.63 | 0.58 |
| Lagged treatment effect | 0.05 | 0.05 | 0.76 | 0.86 | 0.66 | 0.82 | 0.78 | 0.58 | 0.73 |
| Faster recoveries | 0.05 | 0.86 | 0.93 | 0.93 | 0.87 | 0.87 | 0.89 | 0.65 | 0.59 |
| Higher mortality rate | 0.05 | 0.76 | 0.85 | 0.88 | 0.80 | 0.81 | 0.82 | 0.75 | 0.71 |
| Mortality differences only | 0.05 | 0.23 | 0.26 | 0.32 | 0.24 | 0.31 | 0.28 | 0.51 | 0.46 |

Table 4. Evaluation of methods applied to ACTT-1 study data: observed data and simulated sample sizes of 50, 150, and 300 per group. Subsets of data were replicated 100,000 times.

| | | Observed data (n=1059) | | | Empirical power from simulations | | |
|---|---|---|---|---|---|---|---|
| | | Estimates | 95% CI | p-value | 50 per group | 150 per group | 300 per group |
| Proportional odds model | Day 3 | 1.49 | (1.16,1.82) | 0.001 | 0.16 | 0.39 | 0.77 |
| | Day 5 | 1.54 | (1.23,1.93) | <0.001 | 0.20 | 0.52 | 0.91 |
| | Day 7 | 1.62 | (1.29,2.04) | <0.001 | 0.24 | 0.62 | 0.97 |
| | Day 10 | 1.61 | (1.26, 2.04) | <0.001 | 0.21 | 0.55 | 0.94 |
| | Day 14 | 1.50 | (1.18,1.92) | 0.001 | 0.16 | 0.41 | 0.79 |
| | Day 21 | 1.42 | (1.09,1.85) | 0.009 | 0.11 | 0.25 | 0.50 |
| | Day 28 | 1.34 | (0.99,1.82) | 0.063 | 0.07 | 0.13 | 0.19 |
| Mean difference (t-test) | Day 3 | 0.28 | (0.11,0.46) | 0.002 | 0.15 | 0.37 | 0.74 |
| | Day 5 | 0.45 | (0.22,0.68) | <0.001 | 0.20 | 0.52 | 0.91 |
| | Day 7 | 0.56 | (0.29,0.83) | <0.001 | 0.22 | 0.59 | 0.95 |
| | Day 10 | 0.58 | (0.28,0.88) | <0.001 | 0.20 | 0.53 | 0.91 |
| | Day 14 | 0.62 | (0.27, 0.96) | <0.001 | 0.17 | 0.46 | 0.85 |
| | Day 21 | 0.53 | (0.18, 0.87) | 0.003 | 0.13 | 0.33 | 0.67 |
| | Day 28 | 0.41 | (0.07, 0.74) | 0.017 | 0.09 | 0.20 | 0.40 |
| | All-days average | 0.55 | (0.31, 0.79) | <0.001 | 0.27 | 0.69 | 0.98 |
| Time to event (log rank) | Recovery rate ratio | 1.32 | (1.12, 1.55) | <0.001 | 0.18 | 0.48 | 0.87 |
| | Improvement rate ratio (1-point) | 1.29 | (1.11, 1.49) | <0.001 | 0.19 | 0.51 | 0.90 |
| | Improvement rate ratio (2-point) | 1.28 | (1.10, 1.50) | 0.001 | 0.17 | 0.44 | 0.84 |
| | Hazard ratio (death) | 0.70 | (0.47, 1.04) | 0.073 | 0.07 | 0.12 | 0.18 |

Figure 1. Ordinal outcome values by day of study from 100 simulated trajectories from the reference scenario. The smooth lines represent the average trajectories, while the bent lines represent the observed scores for individual patients.

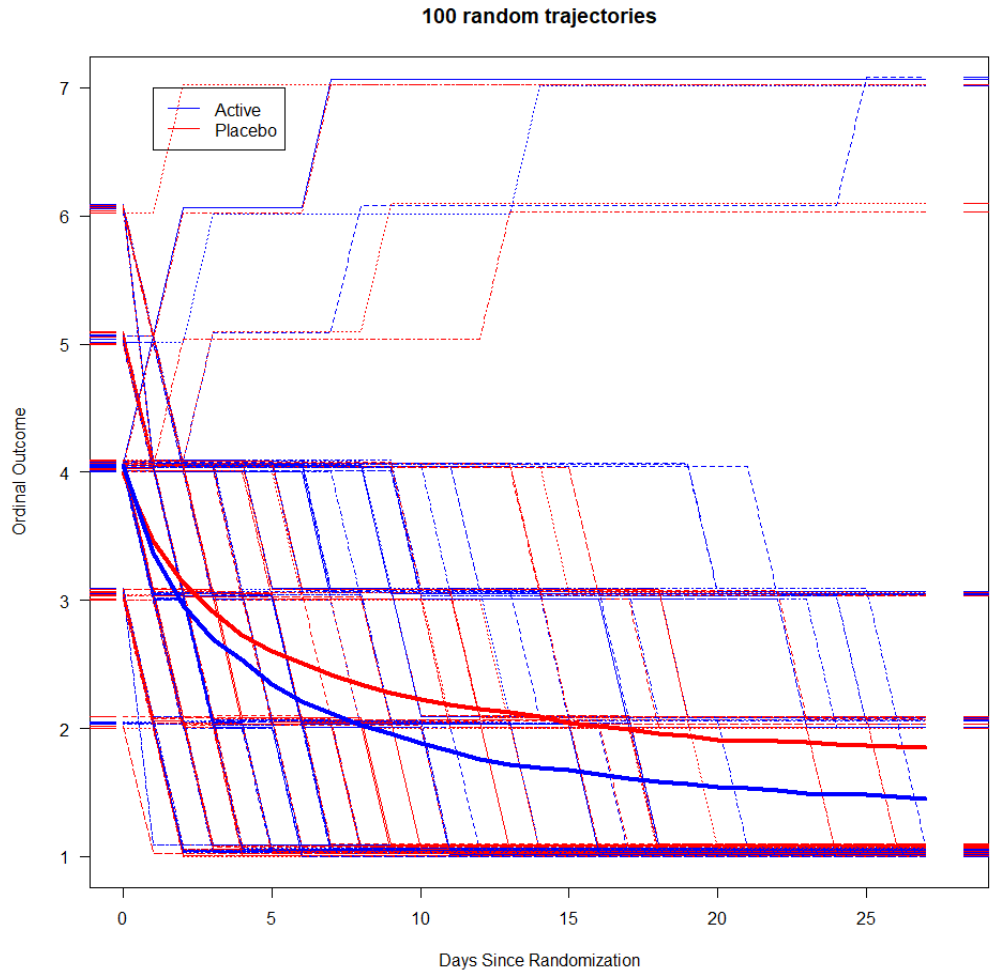

**Funding statement**

*The author(s) disclosed receipt of the following financial support for the research, authorship, and/or publication of this article:*

J Wang and T Bonnett received funds from the National Cancer Institute, National Institutes of Health, under Contract No. 75N91019D00024, Task Order No. 75N91019F00130.  The content of this publication does not necessarily reflect the views or policies of the Department of Health and Human Services, nor does mention of trade names, commercial products, or organizations imply endorsement by the U.S. Government.

T Jaki received funding from UK Medical Research Council (MC_UU_0002/14). This report is independent research arising in part from Prof Jaki's Senior Research Fellowship (NIHR-SRF-2015-08-001) supported by the National Institute for Health Research. The views expressed in  this publication are those of the authors and not necessarily those of the NHS, the National Institute for Health Research or the Department of Health and Social Care (DHCS).

F Koenig and C Schoergenhofer: Medical University Vienna contribution is financially supported by the Austrian Federal Ministry of  Education, Science and Research

**Acknowledgments**

The authors wish to acknowledge the ACTT-1 and LOTUS study teams for use of data from their study.

APPENDIX:

Table S1. Selected Clinical Trials for Covid19 with outcomes and ordinal scales

| Study | Primary outcome | Ordinal scale | Treatments | Study Design | Population |
|---|---|---|---|---|---|
| LOTUS ChiCTR2000029308 | <u>Time to clinical improvement</u><br><br>Clinical improvement defined as two points improvement on a 7-category ordinal scale or discharge from the hospital, whichever came first. | 1) Not hospitalized with resumption of normal activities<br>2) Not hospitalized, but unable to resume normal activities<br>3) Hospitalized, not requiring supplemental oxygen<br>4) Hospitalized, requiring supplemental oxygen<br>5) Hospitalized, requiring nasal high-flow oxygen therapy, noninvasive mechanical ventilation, or both<br>6) Hospitalized, requiring ECMO, invasive mechanical ventilation or both<br>7) Death. | 1) Lopinavir/ritonavir<br>2) Standard of care | Randomized, controlled, open-label trial.<br><br>Randomization ratio: 1:1.<br><br>Final:199<br>99 Lopinavir-Ritonavir<br>100 SOC | **Severe Covid-19 patients** hospitalized adult patients with confirmed SARS-CoV-2 infection, and Sao2 <94% while breathing ambient air or Pao2/Fio2 < 300 mm Hg. |
| ACTT NCT04280705 | <u>Time to recovery 28 days from randomization</u><br><br>Recovery defined as category 1, 2 or 3. | 1) Not hospitalized, no limitations on activities<br>2) Not hospitalized, limitation on activities and/or requiring home oxygen<br>3) Hospitalized, not requiring supplemental oxygen - no longer requires ongoing medical care<br>4) Hospitalized, not requiring supplemental oxygen - requiring | Stage 1:<br>1) Remdesivir<br>2) Placebo<br><br>Stage 2:<br>1) Remdesivir + baricitinib<br>2) Remdesivir | Adaptive randomized, double-blind, placebo-controlled platform trial.<br><br>Randomization ratio: 1:1<br><br>Stage 1 sample size: 400 recoveries<br><br>Stage 2 sample size: 723 recoveries | Hospitalized adults with COVID-19, mild, moderate, and severe patients |

| | | ongoing medical care (COVID-19 related or otherwise)<br>5) Hospitalized, requiring supplemental oxygen<br>6) Hospitalized, on non-invasive ventilation or high flow oxygen devices<br>7) Hospitalized, on invasive mechanical ventilation or ECMO<br>8) Death | | | |
|---|---|---|---|---|---|
| Remdesivir in Adults with Severe COVID-19<br>NCT04257656 | <u>Time to clinical improvement:</u><br><br>Clinical improvement defined as two points improvement on a 6-category ordinal scale or discharge from the hospital, whichever came first. | 1) hospital discharge;<br>2) hospitalized, not requiring supplemental oxygen;<br>3) hospitalized, requiring supplemental oxygen;<br>4) Hospitalized, requiring nasal high-flow oxygen therapy, noninvasive mechanical ventilation, or both<br>6) Hospitalized, requiring ECMO, invasive mechanical ventilation or both<br>6) death; | 1) Remdesivir<br>2) Placebo | Randomized, double-blind, placebo-controlled<br><br>Randomization ratio: 2:1<br><br>Planned sample size: 325 clinical improvements | Adults (≥18 years) with laboratory confirmed COVID-19 virus infection, and severe pneumonia signs or symptoms, and radiologically confirmed severe pneumonia (severe patients) |

| Trial | Primary Outcome | Ordinal Scale | Treatment Arms | Design | Population |
|---|---|---|---|---|---|
| Randomized Evaluation of COVID-19 Therapy (RECOVERY) ISRCTN 50189673 | All-cause mortality at 28 days after first randomization | None | First randomization:<br>1) Lopinavir/ritonavir<br>2) Low-dose Corticosteroid<br>3) Hydroxychloroquine<br>4) Azithromycin<br>5) Standard of care<br><br>Second randomization (in worsening patients):<br>1) Tocilizumab<br>2) Standard of care | Adaptive, randomized, placebo-controlled, multicenter, multi-arm designed, open-label trial<br><br>Planned sample size: unknown, depending on scale of pandemic | Hospitalized adults with SARS-CoV-2 infection (clinically suspected or laboratory confirmed), mild, moderate, and severe patients |
| Trial of Treatments for COVID-19 in Hospitalized Adults (DisCoVeRy) NCT04315948 | Day 15 subject clinical status on 7-point ordinal scale | 1) Not hospitalized, no limitations on activities<br>2) Not hospitalized, limitation on activities<br>3) Hospitalized, not requiring supplemental oxygen<br>4) Hospitalized, requiring supplemental oxygen<br>5) Hospitalized, on non-invasive ventilation or high flow oxygen devices<br>6) Hospitalized, on invasive mechanical ventilation or ECMO<br>7) Death. | 1) Remdesivir<br>2) Lopinavir/ritonavir<br>3) Lopinavir/ritonavir + Interferon ß-1a<br>4) Hydroxychloroquine<br>5) Standard of care | Adaptive, randomized, open-label clinical trial<br><br>Randomization ratio: participants 1:1:1:1:1<br><br>Planned sample size: 3100 participants | Hospitalized adult patients with laboratory-confirmed SARS-CoV-2 infection as determined by PCR, in any specimen < 72 hours prior to randomization<br><br>Clinical assessment of pneumonia (evidence of rales/crackles on exam) AND SpO2 ≤ 94% on room air OR acute respiratory failure requiring supplemental oxygen, high flow oxygen devices, non-invasive ventilation, and/or mechanical ventilation. |
| Austrian CoronaVirus Adaptive | Time to clinical improvement | The 7-categories of the World Health Organization proposed scale, as | 1) Hydroxychloroquine<br>2) Lopinavir/ritonavir<br>3) Standard of care | A multicenter, randomized, open label, controlled platform trial | Laboratory confirmed (i.e. PCR-based assay) infection with SARSCoV- |
| Clinical Trial (ACOVACT) NCT04351724 | defined as time from randomization to a sustained improvement of at least one category on two consecutive days compared to the status at randomization measured on a seven-category ordinal scale (proposed by WHO). | follows:<br>1. Not hospitalized, no limitations on activities<br>2. Not hospitalized, limitation on activities;<br>3. Hospitalized, not requiring supplemental oxygen;<br>4. Hospitalized, requiring supplemental oxygen;<br>5. Hospitalized, on non-invasive ventilation or high flow oxygen devices;<br>6. Hospitalized, on invasive mechanical ventilation or ECMO;<br>7. Death. | 4) Pooled plasma or IVIG from reconvalescent patients*<br><br>*Treatment arm will only be opened when product and the respective necessary documents are available. | Randomization ratio for anti-viral treatment arms 1:1:1<br><br>Planned sample size is 500 participants<br><br>The main study is for the comparison of anti-viral treatments<br><br>Interim analysis after 50 patients in a treatment arm<br><br>ACOVAT includes further sub-studies with additional randomization on top of the anti-viral treatments | 2 (ideally but not necessarily ≤72 hours before randomization for "antiviral" treatments) OR radiological signs of COVID-19 in chest X-ray or computed tomography<br>• Hospitalization due to SARS-CoV-2 infection (for anti-viral treatment arms)<br>• Requirement of oxygen support (due to oxygen saturation <94% on ambient air or >3% drop in case of chronic obstructive lung disease) OR radiological signs of COVID-19 |

Table S2. Statistical analysis strategies including advantages and disadvantages.

| Endpoint | Possible statistical analysis strategy | Advantages | Disadvantages |
|---|---|---|---|
| Binary analyses | | | |
| 1. Proportion recovered/improved (by one or two categories on an ordinal scale) from baseline to specified time point like 2 weeks. | $\chi^2$-Test, Boschloo's test of proportions, logistic regression | Accounts for baseline, clinically relevant, interpretation | Fixed time, loss of power due to dichotomization and using a binary endpoint |
| 2. Mortality by day 28 | $\chi^2$-Test, Boschloo's test of proportions, logistic regression | Clinically meaningful, easy to interpret | Requires large sample sizes when mortality rate low |
| Ordinal scale analyses | | | |
| 3. Ordinal outcome such as a 6-point scale at a fixed time point (e.g., 2 weeks), | Wilcoxon | Captures multiple states | Fixed time, no baseline, ties, scale categories should be objective and clinically meaningful, interpretation |
| 4. Change in ordinal scale from baseline to follow-up | t-test or Wilcoxon | Accounts (partly) for baseline | Fixed time, edge effect (little room for improvement/worsening for those at tails/edges), ties, interpretation |
| 5. Ordinal scale at a fixed time point (e.g., 2 weeks). | Proportional odds model | More robust (no normality assumption), score test is asymptotically like Wilcoxon test | Fixed time, Assumption of constant treatment to control odds ratio for each 1 unit change in ordinal scale; efficiency |
| 6. Ordinal outcome at a fixed time point adjusted for baseline value | Generalized proportional odds model, analysis of covariance (ANCOVA) | Accounts for baseline, power, and is equivalent to the analysis of endpoint (4) when using Change in ordinal outcome from baseline to fixed time point with an ANCOVA adjusting for baseline as covariate | Fixed time, edge effect (little room for improvement/worsening for those at tails/edges), ties, interpretation |
| 7. Average of ordinal scale over daily (or at least frequent) measurements during follow-up. | (potential analysis see 4) | Covers a predefined range of days, power | Duration and severity are mixed (e.g., 1-day death equals 7 days healthy), clinical relevance? Diluted effect if treatment effect established later |
| 8. Average of ordinal scale over daily (or at least frequent) measurements during follow-up minus baseline ordinal scale measurement. | (potential analysis see 4) | covers a predefined range of days, power, accounts for baseline | Duration and severity are mixed, clinical relevance? Diluted effect if treatment effect established later |
| 9. Average of ordinal scale over daily (or at least frequent) measurements during follow-up adjusted for baseline | (potential analysis see 6, ANCOVA) | covers a predefined range of days, more power than 8, accounts for baseline | Duration and severity are mixed, clinical relevance, interpretation, Diluted effect if treatment effect established later |
| 10. Area under the curve of ordinal scale over frequent measurements. | Endpoint similar to 7 (potential analyses see 3 or 7) | covers a predefined range of days, area smaller if less time (similar problem for 7, 8, and 9) | Clinical relevance, diluted effect if treatment effect established later |
| Time-to-event analyses | | | |
| 11. Time to a specified level of improvement (e.g., time to recovery) | Log-rank test (or Cox Regression) Deaths censored Max follow-up | Captures time element Interpretation: rate of recovery and median days to recovery | Does not consider starting point and individual courses to improvement (For unstratified log-rank test) |
| 12. Time to a specific magnitude of improvement (e.g., 2-point improvement in ordinal scale) | Log-rank test (or Cox Regression) Deaths censored Max folllowup | Captures time element Interpretation: rate of 2-point improvement, median days to 2-point improvement | Improvements are considered equally regardless of starting point (e.g., from 6 to 4 considered equal to 3 to 1) (For Cox proportional hazard assumption) |
| 13. Time to recovery and time to death | Standard Kaplan-Meier & Cox for death. Fine-Gray models for recovery | Provides treatment effects on two different aspects | Unclear how to combine the treatment effects in a single analysis. |
| Continuous data analyses | | | |
| 14. Difference in days of oxygen use/intubation/etc | t-test, Wilcoxon | Possible statistical efficiency | May not correlate with eventual outcome |
| 15. Difference in viral loads | t-test, Wilcoxon | Possible statistical efficiency | May not correlate with eventual outcome |
| 16. Various biomarkers | t-test, Wilcoxon | Possible statistical efficiency | May not correlate with eventual outcome |

Table S3. Demonstration of difference in statistical methods applied to reported study data from the LOTUS lopinavir/ritonavir study

|  | Based on observed data (n=199) | Hypothetical example each observation included twice (n=398)† |
|---|---|---|
| **Proportional odds model** | | |
| Day 7 odds ratio | 1.206 (95% CI: 0.710, 2.054) p=0.488 | p=0.327 |
| Day 14 odds ratio | 1.376 (95% CI: -0.835, 2.274) p=0.212 | p=0.077 |
| Day 21 odds ratio | 1.196 (95%CI: 0.676, 2.124) p=0.539 | p=0.386 |
| Day 28 odds ratio | 1.370 (95%CI: -0.740, 2.563) p=0.319 | p=0.159 |
| **Average score (t-test)** | | |
| Mean difference | -0.1678 (95%CI: -0.575;0.240) p=0.418 | p=0.250 |
| **Average score change from baseline (t-test)** | | |
| Mean difference, change from baseline | -0.247 (95%CI: -0.628,0.134) p=0.202 | p=0.070 |
| **Time-to-recovery (log rank test)** | | |
| Recovery rate ratio | 1.248 (95%CI: 0.899,1.732) p=0.187 | p=0.061 |
| **Time-to-improvement (log rank test)** | | |
| Beneficial ratio | 1.307 (95%CI: 0.946;1.807) p=0.105 | p=0.022 |
| **Mortality (Fisher's exact test)** | | |
| Odds ratio | 0.786 (95%CI:0.372, 1.644) p=0.602 | p=0.390 |

Details of simulation

Ordinal trajectories for each subject were generated according to a linear random effects model with time index log of the day since randomization  Informally, subject i drew a random curve  of 'destiny' and foreach day of follow-up, the integer part of the line at that day was given as ordinal score.  Except for the lagged effect scenario, the model is given by

$$Y_{id} = B_0 + B_1 \log(d) + B_2 Z \log(d) + b_{0i} + b_{1i} * \log(d) + W\, e_{id} \qquad (1)$$

with $b_{0i}$ distributed $N(0, 1.5^2)$ and $b_{1i}$ distributed $I \times N(-4, .3^2) + (1-I)\, N(7, s^2)$ with I distributed Bernoulli(p=0.10) for placebo and Bernoulli(p=0.05)~Be(.05) for treatment, $e_{id}$ distributed $N(0, .25^2)$, and Z the indicator of the treatment group.   Note that there is a treatment effect both on the speed of recovery (as B2 <0) and mortality as I has a different Bernoulli probability for the two groups.

For the lagged effect scenario, the day 1 treatment effect begins at day 8:

$$Y_{id} = B_0 + B_1 \log(d) + B_2 Z\, I(d>7)*\log(d-7) + b_{0i} + b_{1i} *Z*I(d>7) \log(d-7) + W\, e_{id} \qquad (2)$$

With settings for the random variables as described for equation (1). Table S4 provides the parameter values used for the different scenarios.

**Table S4:** Table of parameters used for the various model.  The feature that is changed relative to the reference case is bolded.   All scenarios use equation (1) except for the lagged effect which uses equation (2)

| Scenario | B0 | B1 | B2 | s | W |
|---|---|---|---|---|---|
| Reference | 0 | -.05 | -.10 | .15 | 0 |
| **Lagged Effect*** | 0 | -.05 | -.10 | .15 | 0 |
| Faster Recovery | 0 | **-.10** | -.10 | .15 | 0 |
| Faster Mortality | 0 | -.05 | -.10 | **.30** | 0 |
| Only Mortality benefit | 0 | -.05 | **0** | .15 | 0 |

Details of simulation enforcing proportional odds assumption at each time point

Random multinomial data were generated corresponding to baseline ordinal scores. Then a trajectory of ordinal scores was applied as method 1 above, except that the trajectories were generated with the same distribution for treatment and control arms.  Treatment-arm proportions at observation days were then re-scaled to satisfy a proportional odds assumption with according to a common odds ratio for specific treatment effects each day (as specified in table S5). Additional simulation studies (not shown) demonstrated that blinded (pooled) pilot studies are not very informative for guiding the determination of the optimal time.  Blinded (pooled) data provide information about the overall proportions in each category, but simple rules such as selecting the time where there are a certain proportion of good outcomes or when the distribution is the most variable do not seem to improve identification of the optimal time for evaluation.  Note the one peculiarity of how these models are set up.

**Table S5. Simulated power for different tests under different scenarios.**

|  | True common odds ratio by day | | | | | Empirical Power/Rejection Rates | | | |
| --- | --- | --- | --- | --- | --- | --- | --- | --- | --- |
|  | Days 1-10 | Days 11-13 | Day 14 | Day 21 | Day 28 | Proportional odds at day 14 | Proportional odds at day 28 | Log-rank (time to recovery) | Average score |
| Scenario A | 1 | 1 | 1 | 1.5 | 1.75 | 0.052 | 0.879 | 0.395 | 0.271 |
| Scenario B | 1 | 1 | 1.25 | 1.5 | 1.75 | 0.244 | 0.884 | 0.442 | 0.384 |
| Scenario C | 1 | 1.1 | 1.15 | 1.25 | 1.75 | 0.126 | 0.884 | 0.418 | 0.254 |

**Figure S1. Stacked bar plots for ordinal scores and Kaplan-Meier curves for time-to-recovery for three scenarios for simulation method 2.**

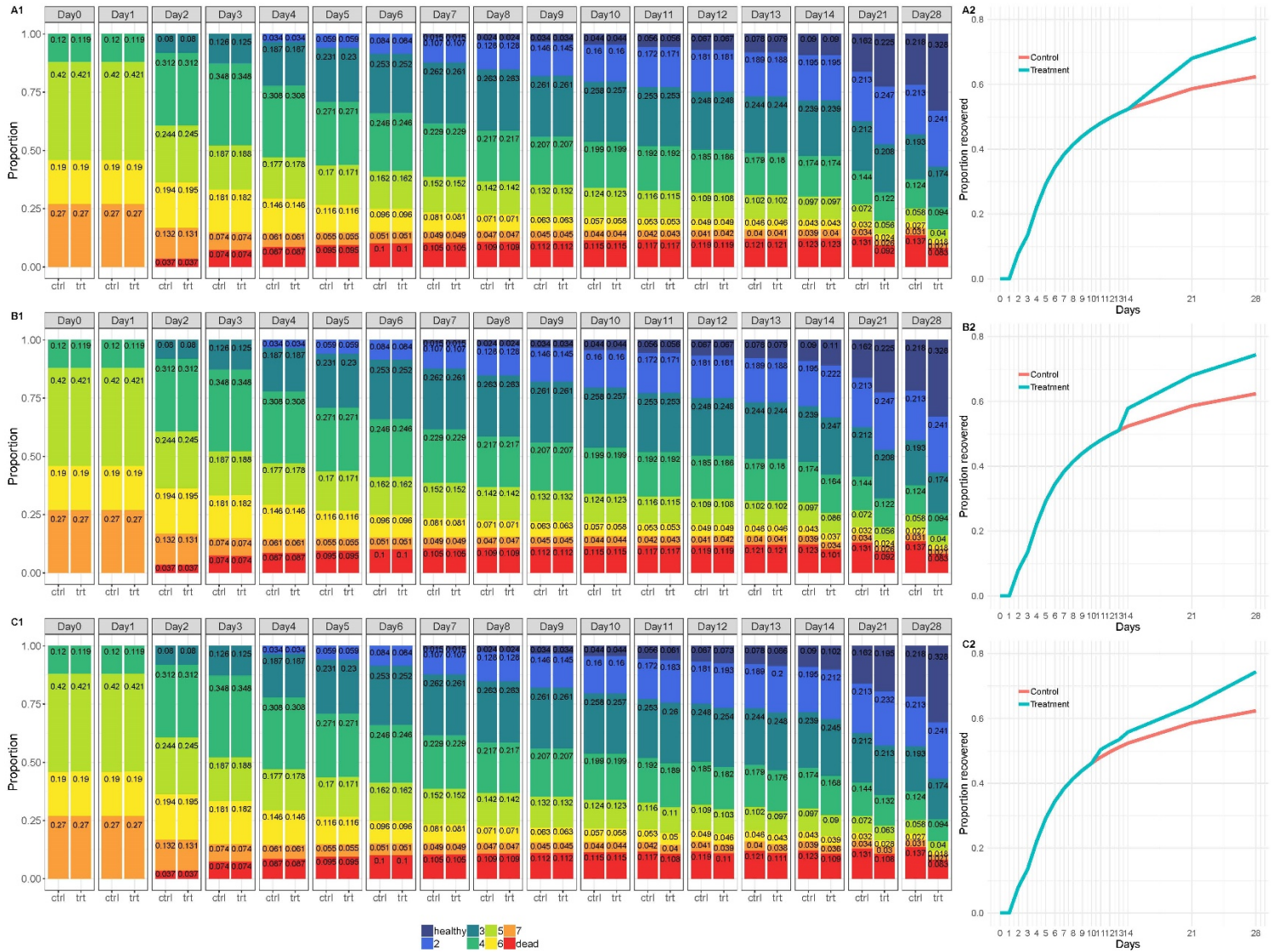